\documentclass[
 aip,apl,
 amsmath,amssymb,
reprint]{revtex4-1}

\usepackage{graphicx}
\usepackage{dcolumn}
\usepackage{bm}
\usepackage{color}
\usepackage{url}
\usepackage{longtable}

\begin{document}
\title{Near atomically smooth alkali antimonide photocathode thin films}

\author{Jun Feng}
\email{fjun@lbl.gov}
\affiliation{Lawrence Berkeley National Laboratory, Berkeley, CA, USA, 94720}

\author{Siddharth Karkare}
\affiliation{Lawrence Berkeley National Laboratory, Berkeley, CA, USA, 94720}
\author{James Nasiatka}
\affiliation{Lawrence Berkeley National Laboratory, Berkeley, CA, USA, 94720}
\author{Susanne Schubert}
\affiliation{Lawrence Berkeley National Laboratory, Berkeley, CA, USA, 94720}

\author{John Smedley}
\affiliation{Brookhaven National Laboratory, Upton, NY, USA, 11973}

\author{Howard Padmore}
\affiliation{Lawrence Berkeley National Laboratory, Berkeley, CA, USA, 94720}

\begin{abstract}
Nano-roughness limits the emittance of electron beams that can be generated by high efficiency photocathodes, such as the thermally reacted alkali antimonide thin films. However there is an urgent need for photocathodes that can produce an order of magnitude or more lower emittance than present day systems in order to increase the transverse coherence width of the electron beam. In this paper we demonstrate a method for producing alkali antimonide cathodes with near atomic smoothness with high reproducibility.
\end{abstract}

\maketitle

Photoemission based electron sources for the next generation x-ray high repetition rate, high brightness light sources such as Energy Recovery Linacs\cite{ERL} and Free Electron Lasers\cite{FEL1} need to satisfy several criteria, namely: high ($>$1\%) quantum efficiency (QE) in the visible range, smallest possible intrinsic emittance, fast (sub-ps) response time and a long operational lifetime.  
During the past decade, alkali-antimonides (eg. K$_2$CsSb) have emerged as the only class of materials that satisfies all these requirements with a high QE $>$5\% and a low intrinsic emittance in the range of 0.36-0.5 $\mu$m per mm rms laser spot size in green (520-545 nm) light  \cite{sol_tech,theo_KCsSb,highcurr}. Additionally, alkali-antimonides also show promise as sources of ultra-cold electrons for ultrafast electron diffraction\cite{UED} applications and Inverse Compton Scattering based Gamma ray sources\cite{IC}. 


Although alkali antimonides have many excellent characteristics, the synthesis process leads to relatively high levels of roughness \cite{KCsSb_APLM2}.
K$_2$CsSb photocathodes are typically grown as thin films over conducting substrates by thermal evaporation of $\sim$10-30 nm of Sb followed by sequential thermal evaporation and reaction of K and Cs respectively \cite{KCsSb_APLM,sol_tech}. The films created by this process are not ordered and can have a root mean square (rms) surface roughness as high as 25 nm with a period of roughly 100 nm\cite{KCsSb_APLM2}. Such a surface roughness can distort the electric field near the cathode surface causing the intrinsic emittance to drastically increase. Ignoring the contribution of the slope effect\cite{karkare,bradley}, to first order,  the intrinsic emittance after accounting for this electric field effect can be given by $\epsilon_{in}=\sqrt{\epsilon_{in0}^2+\epsilon_f^2}$, where $\epsilon_{in0}$ is the intrinsic emittance of the cathode at near zero electric field and $\epsilon_f$ is the enhancement to the intrinsic emittance at an electric field of $f$ MV/m (typically in the range of 1-20 MV/m) at the cathode surface.
In RF/SRF based electron guns, used for high bunch charge applications, the electric field at the cathode surface can be greater than 20 MV/m. In this case the electric field enhancement of the intrinsic emittance can be as high as 2 $\mu$m per mm rms laser spot size making these cathodes unusable.\cite{JOHN2015_IPAC}. 

The smallest possible intrinsic emittance is limited by the lattice temperature of the cathode to $\epsilon_{in0}=0.22$ $\mu$m at room temperature and can be obtained by exciting electrons with near threshold photons\cite{jun_thermal}. However, in alkali-antimonides the smallest possible emittance is limited to a higher value even at photoemission threshold because of the surface roughness\cite{cold_prstab}. In order to reach the thermal limit and attain the smallest possible intrinsic emittance from cryo-cooled alkali antimonide cathodes a sub-nm surface roughness would be required\cite{cold_prstab}.  

Apart from the surface roughness, another drawback of the traditional sequential growth procedure is its irreproducibility and unreliability. Despite the wide use of this growth technique for streak camera and photomultiplier applications since the 1960s\cite{sommer,alkali_app}, this growth technique is complex and remains extremely sensitive to many deposition parameters such as substrate and source temperature, growth rate, final thickness and the quality of vacuum, etc. making it difficult to control and reproduce\cite{dowell_alkali}. The complexity of this technique makes it difficult to automate and the results depend strongly on the skill of the cathode grower\cite{nakamura}.

Recently, X-ray diffraction studies have shown that this surface roughness and the irreproducibility in this traditional growth process can be attributed to the exothermic reaction of K (or other alkali metals) deposited on top of the Sb thin film\cite{KCsSb_APLM,KCsSb_APLM2,susanne_PRSTAB}. This reaction leads to several meta-stable K-Sb states making the process difficult to control.

In this paper, we report a growth procedure to grow K-Cs-Sb cathodes using a co-deposition technique. This technique avoids the exothermic reaction between a previously deposited Sb film and alkali metals making it very reproducible and is less complicated than the traditional sequential deposition technique. We present the spectral response and atomic force microscopy (AFM) measurements of the K-Cs-Sb cathodes grown using this technique and show that they are similar in QE, but are far smoother as measured by AFM when compared to the cathodes grown using the traditional sequential deposition technique. The typical roughness of cathodes grown using the co-deposition technique is less than 1 nm (rms). This technique will not only allow easy production of alkali antimonide photocathodes, but will also allow their use in high electric field RF/SRF guns and enable generation of ultra-cold electrons from these cathodes.

\begin{figure}
\includegraphics[scale=0.3]{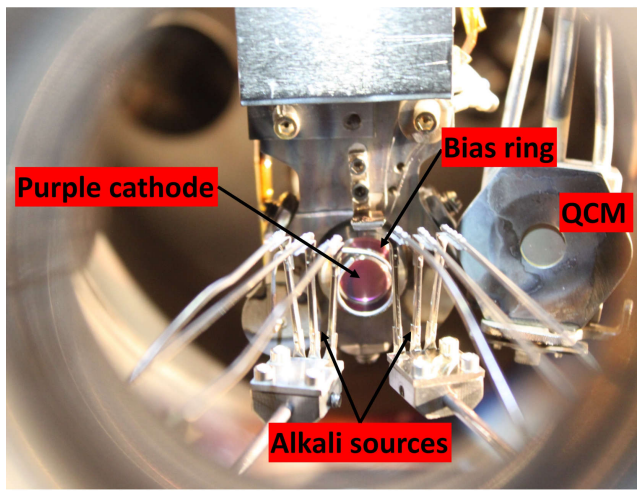}
\caption{Picture of the co-deposited cathode along with the alkali metal sources mounted in the UHV deposition chamber\label{fig1}}
\end{figure}

Several photocathodes were prepared on Mo and Si substrates, 12 mm in diameter. The surface of the substrates was optically polished and heated to 400$^{\circ}$C for 30 minutes in a UHV deposition chamber for cleaning prior to growth. The base pressure in the UHV deposition chamber was 1$\times$10$^{-10}$ torr. Sb is evaporated by heating 99.9999\% pure Sb pellets (obtained from Alfa Aesar\cite{alfaaesar}) in a ceramic furnace. A pre-calibrated quartz crystal microbalance (QCM) is used to measure the Sb deposition rate. The alkali metals are simultaneously evaporated from 17 mm getter sources manufactured by SAES\cite{SAES}. The alkali metal sources are mounted 30 mm away from the substrate as shown in figure \ref{fig1}.
The substrate is grounded and a bias ring (biased to 50 V) is used to collect the photocurrent emitted from the cathode during and after deposition to measure the QE.

The substrate temperature is maintained at 90$^{\circ}$C during the entire deposition process. Sb, K and Cs are deposited simultaneously on the substrate. Sb is deposited at a rate of 0.1\AA/s as measured by the QCM. This assumes a sticking coefficient of 1, as demonstrated by cross calibration with x-ray reflectivity and scanning electron microscopy measurements\cite{JOHN2012_IPAC}.  The alkali metals are evaporated by passing currents of 4.7 A and 5.5 A through the K and  Cs sources respectively. No parameter needs to be varied during the entire process.  The deposition rate of K and Cs cannot be similarly measured using a QCM due to the unknown sticking coefficients of the alkali metals on the QCM and variability caused by temperature and deposition history\cite{alkali_sticking}.

\begin{figure}
\includegraphics[scale=.3]{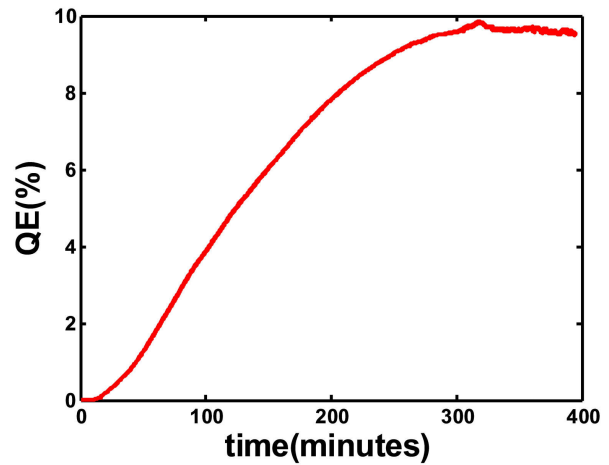}
\caption{QE of the K-Cs-Sb cathode grown using co-deposition process as a function of growth time\label{fig2}}
\end{figure}

A 532 nm laser is used to monitor the QE during deposition. Figure \ref{fig2} shows the QE as a function of deposition time. Evaporation of all the three metals is started simultaneously. The deposition rate is held constant throughout the deposition process. As seen from figure \ref{fig2}, the QE starts to increase about 9 minutes after starting the deposition process and continues to increase up to 8-10\% for roughly 300 mins. At this point the QE reaches a maxima and then saturates. All the sources are turned off at this point and the substrate is allowed to cool. The QE stays nearly constant as the substrate cools down slowly to room temperature. Ten cathodes were grown using this procedure and all gave a final QE between 7-10\%. No significant substrate dependence was found.

\begin{figure}
\includegraphics[scale=.3]{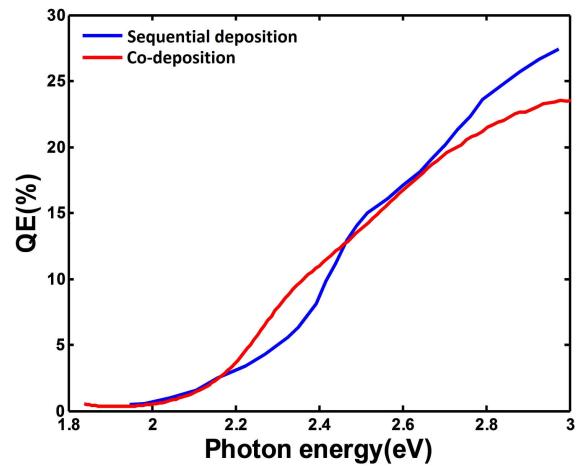}
\caption{Comparison of the spectral response of the K-Cs-Sb cathodes grown by co-deposition and sequential deposition techniques\label{fig3}}
\end{figure}  

The spectral response of the K-Cs-Sb cathode is measured using a plasma based wavelength tunable light source\cite{UVmono}. Figure \ref{fig3} shows a comparison of the K-Cs-Sb cathodes grown using the co-deposition process and the traditional sequential growth process\cite{theo_KCsSb}. We can see that cathodes grown using both techniques have nearly the same photoemission threshold, but the QE of the co-deposition cathode is higher between photon energies of 2.2 eV and 2.4 eV and is lower at photon energies greater than 2.6 eV. These differences can be attributed to possible structural and compositional differences between the cathodes obtained by the two deposition techniques. 

\begin{figure}
\includegraphics[scale=.35]{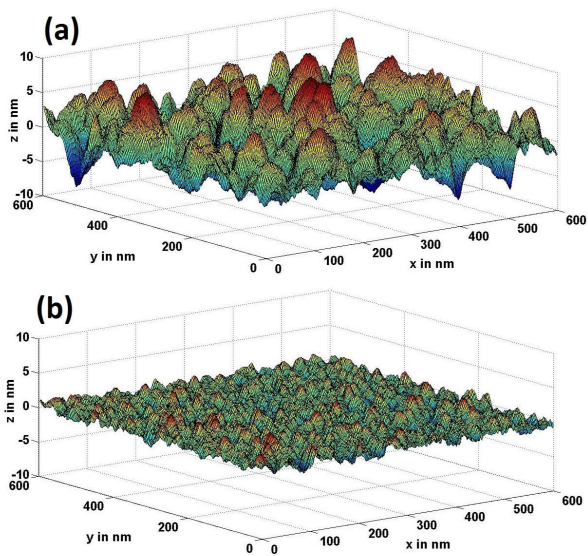}
\caption{(a) AFM image of a cathode grown using sequential deposition (b) AFM image of a cathode grown using the co-deposition technique\label{fig4_afm}}
\end{figure}

In order to compare the surface roughness, two cathodes were grown - one using the co-deposition technique described above and the other using a sequential growth technique\cite{theo_KCsSb} on commercially bought Si substrates with roughness below 0.3 nm. These cathodes were then transported in vacuum into an UHV-AFM. The thickness of the initial Sb layer for the sequential deposition was 7 nm.
Figure \ref{fig4_afm} shows the AFM image of the surface of these cathodes. The rms values of the roughness for the two surfaces are given in table \ref{tab1}.  

The surface can be expanded in its 2D Fourier components and can be written as $z =\sum_{n}A_n\phi_n\left(x,y\right)$. The electric potential near the surface can be given by $U =E_0z+\sum_{n}C_ne^{−z/p_n}\phi_n$, where $p_n^{-1}= \sqrt{p^{-2}_{nx}+p^{-2}_{ny}}$, $p_{nx}$ and $p_{ny}$ are periods corresponding to the Fourier components in x and y directions respectively, $E_0$ is the longitudinal electric field away from the surface and $C_n$ are coefficients obtained by solving the Poisson equation\cite{gorlov}.  If $p_n\gg A_n$ then $C_n\approx A_nE_0$\cite{bradley}. Assuming the initial velocity of the electron to be 0 and the transverse position to be nearly constant during the acceleration one can calculate the final velocity in the $x$ direction due to the electric field as $v_x=\frac{e}{2m_e}\sum_nC_n\frac{d\phi_n}{dx}\sqrt{\frac{2\pi m_ep_n}{eE_0}}$. From this one can easily calculate the enhancement in intrinsic emittance (per unit laser spot size) as $\epsilon_f=\sqrt{\frac{e\pi^2}{2m_ec^2E_0}\sum_nC_n^2p_n/p_{nx}^2}$. Since $C_n$ is directly proporional to $E_0$, $\epsilon_f\propto\sqrt{E_0}$. It was found that 20 harmonics in the Fourier expansions of the measured surfaces are sufficient to model the surfaces accurately. Hence the values of $p_{nx}$ and $p_{ny}$ are 600/K nm, where K ranges from 0 to 20.

The enhancement in intrinsic emittance due to the distorted electric field calculated using the above formalism for the two surfaces due to surface electric fields of 5 MV/m ($\epsilon_5$) and 20 MV/m ($\epsilon_{20}$) is shown in table \ref{tab1}. 
We can see from table \ref{tab1} that the emittance generated from the rough sequentially evaporated cathode (0.36 $\mu$m/mm rms) at an accelerating field of 20 MV / m is higher than the intrinsic room temperature minimum emittance (0.22 $\mu$m/mm rms). 20 MV/m corresponds to the electric field in a VHF RF photogun\cite{APEX}. Far higher fields, up to 100 MV/m can be generated in high frequency RF guns, and these would make the emittance degradation far worse. On the other hand, the co-deposited films have a much smaller effect on emittance, leading to an increase in emittance of only 0.12 $\mu$m/mm rms. This is significantly less than the minimum room temperature emittance. The thermal and roughness contribution to the emittance in this case should become equal at around 90 K.

In conclusion we have demonstrated a co-deposition based technique to grow ultrasmooth alkali-antimonide cathodes. The technique is significantly simpler than the traditional sequential methodology, easily reproducible and can lead to the automation of the growth process of alkali antimonide cathode. Additionally, this technique produces cathodes with sub-nm scale roughness allowing their use in high bunch charge applications where electric fields at the cathode exceed 20 MV/m without significant degradation in intrinsic emittance. In theory, the surface roughness is small enough to allow production of ultra-cold electron beams from alkali-antimonide cathodes enabling futuristic applications of these cathodes. However, in practice, production of ultra-cold electron may still be limited by other effects like work function non-uniformities and surface defects\cite{nonuniformity}. 
The stoichiometry and structure of the material obtained by this technique is still under investigation.

\begin{table}
\begin{tabular}{ | l | c | c |}
    \hline
    Quantity & Sequential deposition & Co-deposition \\ \hline
    RMS roughness (nm) & 2.5 & 0.6 \\ \hline 
    $\epsilon_{5}$ ($\mu$m)  & 0.18 & 0.06 \\    \hline
    $\epsilon_{20}$ ($\mu$m) & 0.36 & 0.12 \\    \hline
  \end{tabular}
    \caption{RMS roughness and the calculated enhancement in emittance for electric fields of 5 MV/m and 20 MV/m for the two surfaces shown in figure \ref{fig4_afm}\label{tab1}}
\end{table}

This work was performed at LBNL under the auspices of the Office of Science, Office of Basic Energy Sciences, of the U.S. Department of Energy under Contract No. DE-AC02-05CH11231, KC0407-ALSJNT-I0013, and DE-SC0005713

\providecommand{\noopsort}[1]{}\providecommand{\singleletter}[1]{#1}


\begin{thebibliography}{27}%
\makeatletter
\providecommand \@ifxundefined [1]{%
 \@ifx{#1\undefined}
}%
\providecommand \@ifnum [1]{%
 \ifnum #1\expandafter \@firstoftwo
 \else \expandafter \@secondoftwo
 \fi
}%
\providecommand \@ifx [1]{%
 \ifx #1\expandafter \@firstoftwo
 \else \expandafter \@secondoftwo
 \fi
}%
\providecommand \natexlab [1]{#1}%
\providecommand \enquote  [1]{``#1''}%
\providecommand \bibnamefont  [1]{#1}%
\providecommand \bibfnamefont [1]{#1}%
\providecommand \citenamefont [1]{#1}%
\providecommand \href@noop [0]{\@secondoftwo}%
\providecommand \href [0]{\begingroup \@sanitize@url \@href}%
\providecommand \@href[1]{\@@startlink{#1}\@@href}%
\providecommand \@@href[1]{\endgroup#1\@@endlink}%
\providecommand \@sanitize@url [0]{\catcode `\\12\catcode `\$12\catcode
  `\&12\catcode `\#12\catcode `\^12\catcode `\_12\catcode `\%12\relax}%
\providecommand \@@startlink[1]{}%
\providecommand \@@endlink[0]{}%
\providecommand \url  [0]{\begingroup\@sanitize@url \@url }%
\providecommand \@url [1]{\endgroup\@href {#1}{\urlprefix }}%
\providecommand \urlprefix  [0]{URL }%
\providecommand \Eprint [0]{\href }%
\providecommand \doibase [0]{http://dx.doi.org/}%
\providecommand \selectlanguage [0]{\@gobble}%
\providecommand \bibinfo  [0]{\@secondoftwo}%
\providecommand \bibfield  [0]{\@secondoftwo}%
\providecommand \translation [1]{[#1]}%
\providecommand \BibitemOpen [0]{}%
\providecommand \bibitemStop [0]{}%
\providecommand \bibitemNoStop [0]{.\EOS\space}%
\providecommand \EOS [0]{\spacefactor3000\relax}%
\providecommand \BibitemShut  [1]{\csname bibitem#1\endcsname}%
\let\auto@bib@innerbib\@empty
\bibitem [{\citenamefont {Gruner}\ \emph {et~al.}(2002)\citenamefont {Gruner},
  \citenamefont {Bilderback}, \citenamefont {Bazarov}, \citenamefont
  {Finkelstein}, \citenamefont {Krafft}, \citenamefont {Merminga},
  \citenamefont {Padamsee}, \citenamefont {Shen}, \citenamefont {Sinclair},\
  and\ \citenamefont {Tigner}}]{ERL}%
  \BibitemOpen
  \bibfield  {author} {\bibinfo {author} {\bibfnamefont {S.~M.}\ \bibnamefont
  {Gruner}}, \bibinfo {author} {\bibfnamefont {D.}~\bibnamefont {Bilderback}},
  \bibinfo {author} {\bibfnamefont {I.}~\bibnamefont {Bazarov}}, \bibinfo
  {author} {\bibfnamefont {K.}~\bibnamefont {Finkelstein}}, \bibinfo {author}
  {\bibfnamefont {G.}~\bibnamefont {Krafft}}, \bibinfo {author} {\bibfnamefont
  {L.}~\bibnamefont {Merminga}}, \bibinfo {author} {\bibfnamefont
  {H.}~\bibnamefont {Padamsee}}, \bibinfo {author} {\bibfnamefont
  {Q.}~\bibnamefont {Shen}}, \bibinfo {author} {\bibfnamefont {C.}~\bibnamefont
  {Sinclair}}, \ and\ \bibinfo {author} {\bibfnamefont {M.}~\bibnamefont
  {Tigner}},\ }\href@noop {} {\bibfield  {journal} {\bibinfo  {journal} {Rev.
  Sci. Instr.}\ }\textbf {\bibinfo {volume} {73}},\ \bibinfo {pages} {1402}
  (\bibinfo {year} {2002})}\BibitemShut {NoStop}%
\bibitem [{\citenamefont {Emma}\ \emph {et~al.}(2010)\citenamefont {Emma},
  \citenamefont {Akre}, \citenamefont {Arthur}, \citenamefont {Bionta},
  \citenamefont {Bostedt}, \citenamefont {Bozek}, \citenamefont {Brachmann},
  \citenamefont {Bucksbaum}, \citenamefont {Coffee}, \citenamefont {Decker}
  \emph {et~al.}}]{FEL1}%
  \BibitemOpen
  \bibfield  {author} {\bibinfo {author} {\bibfnamefont {P.}~\bibnamefont
  {Emma}}, \bibinfo {author} {\bibfnamefont {R.}~\bibnamefont {Akre}}, \bibinfo
  {author} {\bibfnamefont {J.}~\bibnamefont {Arthur}}, \bibinfo {author}
  {\bibfnamefont {R.}~\bibnamefont {Bionta}}, \bibinfo {author} {\bibfnamefont
  {C.}~\bibnamefont {Bostedt}}, \bibinfo {author} {\bibfnamefont
  {J.}~\bibnamefont {Bozek}}, \bibinfo {author} {\bibfnamefont
  {A.}~\bibnamefont {Brachmann}}, \bibinfo {author} {\bibfnamefont {P.~H.}\
  \bibnamefont {Bucksbaum}}, \bibinfo {author} {\bibfnamefont {R.}~\bibnamefont
  {Coffee}}, \bibinfo {author} {\bibfnamefont {F.~J.}\ \bibnamefont {Decker}},
  \emph {et~al.},\ }\href@noop {} {\bibfield  {journal} {\bibinfo  {journal}
  {Nat. Photonics}\ }\textbf {\bibinfo {volume} {4}},\ \bibinfo {pages} {641}
  (\bibinfo {year} {2010})}\BibitemShut {NoStop}%
\bibitem [{\citenamefont {Bazarov}\ \emph {et~al.}(2011)\citenamefont
  {Bazarov}, \citenamefont {Cultrera}, \citenamefont {Bartnik}, \citenamefont
  {Dunham}, \citenamefont {Karkare}, \citenamefont {Li}, \citenamefont {Liu},
  \citenamefont {Maxson},\ and\ \citenamefont {Roussel}}]{sol_tech}%
  \BibitemOpen
  \bibfield  {author} {\bibinfo {author} {\bibfnamefont {I.}~\bibnamefont
  {Bazarov}}, \bibinfo {author} {\bibfnamefont {L.}~\bibnamefont {Cultrera}},
  \bibinfo {author} {\bibfnamefont {A.}~\bibnamefont {Bartnik}}, \bibinfo
  {author} {\bibfnamefont {B.}~\bibnamefont {Dunham}}, \bibinfo {author}
  {\bibfnamefont {S.}~\bibnamefont {Karkare}}, \bibinfo {author} {\bibfnamefont
  {Y.}~\bibnamefont {Li}}, \bibinfo {author} {\bibfnamefont {X.}~\bibnamefont
  {Liu}}, \bibinfo {author} {\bibfnamefont {J.}~\bibnamefont {Maxson}}, \ and\
  \bibinfo {author} {\bibfnamefont {W.}~\bibnamefont {Roussel}},\ }\href@noop
  {} {\bibfield  {journal} {\bibinfo  {journal} {Appl. Phys. Lett.}\ }\textbf
  {\bibinfo {volume} {98}},\ \bibinfo {pages} {224101} (\bibinfo {year}
  {2011})}\BibitemShut {NoStop}%
\bibitem [{\citenamefont {Vecchione}\ \emph {et~al.}(2011)\citenamefont
  {Vecchione}, \citenamefont {Ben-Zvi}, \citenamefont {Dowell}, \citenamefont
  {Feng}, \citenamefont {Rao}, \citenamefont {Smedley}, \citenamefont {Wan},\
  and\ \citenamefont {Padmore}}]{theo_KCsSb}%
  \BibitemOpen
  \bibfield  {author} {\bibinfo {author} {\bibfnamefont {T.}~\bibnamefont
  {Vecchione}}, \bibinfo {author} {\bibfnamefont {I.}~\bibnamefont {Ben-Zvi}},
  \bibinfo {author} {\bibfnamefont {D.~H.}\ \bibnamefont {Dowell}}, \bibinfo
  {author} {\bibfnamefont {J.}~\bibnamefont {Feng}}, \bibinfo {author}
  {\bibfnamefont {T.}~\bibnamefont {Rao}}, \bibinfo {author} {\bibfnamefont
  {J.}~\bibnamefont {Smedley}}, \bibinfo {author} {\bibfnamefont
  {W.}~\bibnamefont {Wan}}, \ and\ \bibinfo {author} {\bibfnamefont {H.~A.}\
  \bibnamefont {Padmore}},\ }\href@noop {} {\bibfield  {journal} {\bibinfo
  {journal} {Appl. Phys. Lett.}\ }\textbf {\bibinfo {volume} {99}},\ \bibinfo
  {pages} {034103} (\bibinfo {year} {2011})}\BibitemShut {NoStop}%
\bibitem [{\citenamefont {Dunham}\ \emph {et~al.}(2013)\citenamefont {Dunham},
  \citenamefont {Barley}, \citenamefont {Bartnik}, \citenamefont {Bazarov},
  \citenamefont {Cultrera}, \citenamefont {Dobbins}, \citenamefont
  {Hoffstaetter}, \citenamefont {Johnson}, \citenamefont {Kaplan},
  \citenamefont {Karkare} \emph {et~al.}}]{highcurr}%
  \BibitemOpen
  \bibfield  {author} {\bibinfo {author} {\bibfnamefont {B.}~\bibnamefont
  {Dunham}}, \bibinfo {author} {\bibfnamefont {J.}~\bibnamefont {Barley}},
  \bibinfo {author} {\bibfnamefont {A.}~\bibnamefont {Bartnik}}, \bibinfo
  {author} {\bibfnamefont {I.}~\bibnamefont {Bazarov}}, \bibinfo {author}
  {\bibfnamefont {L.}~\bibnamefont {Cultrera}}, \bibinfo {author}
  {\bibfnamefont {J.}~\bibnamefont {Dobbins}}, \bibinfo {author} {\bibfnamefont
  {G.}~\bibnamefont {Hoffstaetter}}, \bibinfo {author} {\bibfnamefont
  {B.}~\bibnamefont {Johnson}}, \bibinfo {author} {\bibfnamefont
  {R.}~\bibnamefont {Kaplan}}, \bibinfo {author} {\bibfnamefont
  {S.}~\bibnamefont {Karkare}},  \emph {et~al.},\ }\href@noop {} {\bibfield
  {journal} {\bibinfo  {journal} {Appl. Phys. Lett.}\ }\textbf {\bibinfo
  {volume} {102}},\ \bibinfo {pages} {034105} (\bibinfo {year}
  {2013})}\BibitemShut {NoStop}%
\bibitem [{\citenamefont {Musumeci}\ \emph {et~al.}(2010)\citenamefont
  {Musumeci}, \citenamefont {Moody}, \citenamefont {Scoby}, \citenamefont
  {Gutierrez}, \citenamefont {Bender},\ and\ \citenamefont {Wilcox}}]{UED}%
  \BibitemOpen
  \bibfield  {author} {\bibinfo {author} {\bibfnamefont {P.}~\bibnamefont
  {Musumeci}}, \bibinfo {author} {\bibfnamefont {J.~T.}\ \bibnamefont {Moody}},
  \bibinfo {author} {\bibfnamefont {C.~M.}\ \bibnamefont {Scoby}}, \bibinfo
  {author} {\bibfnamefont {M.~S.}\ \bibnamefont {Gutierrez}}, \bibinfo {author}
  {\bibfnamefont {H.~A.}\ \bibnamefont {Bender}}, \ and\ \bibinfo {author}
  {\bibfnamefont {N.~S.}\ \bibnamefont {Wilcox}},\ }\href@noop {} {\bibfield
  {journal} {\bibinfo  {journal} {Rev. Sci. Instr.}\ }\textbf {\bibinfo
  {volume} {81}},\ \bibinfo {pages} {013306} (\bibinfo {year}
  {2010})}\BibitemShut {NoStop}%
\bibitem [{\citenamefont {Boucher}\ \emph {et~al.}(2009)\citenamefont
  {Boucher}, \citenamefont {Frigola}, \citenamefont {Murokh}, \citenamefont
  {Ruelas}, \citenamefont {Jovanovic}, \citenamefont {Rosenzweig},\ and\
  \citenamefont {Travish}}]{IC}%
  \BibitemOpen
  \bibfield  {author} {\bibinfo {author} {\bibfnamefont {S.}~\bibnamefont
  {Boucher}}, \bibinfo {author} {\bibfnamefont {P.}~\bibnamefont {Frigola}},
  \bibinfo {author} {\bibfnamefont {A.}~\bibnamefont {Murokh}}, \bibinfo
  {author} {\bibfnamefont {M.}~\bibnamefont {Ruelas}}, \bibinfo {author}
  {\bibfnamefont {I.}~\bibnamefont {Jovanovic}}, \bibinfo {author}
  {\bibfnamefont {J.}~\bibnamefont {Rosenzweig}}, \ and\ \bibinfo {author}
  {\bibfnamefont {G.}~\bibnamefont {Travish}},\ }\href@noop {} {\bibfield
  {journal} {\bibinfo  {journal} {Nucl. Instr. Meth. A}\ }\textbf {\bibinfo
  {volume} {608}},\ \bibinfo {pages} {S54} (\bibinfo {year}
  {2009})}\BibitemShut {NoStop}%
\bibitem [{\citenamefont {Schubert}\ \emph {et~al.}(2013)\citenamefont
  {Schubert}, \citenamefont {Ruiz-Oses}, \citenamefont {Ben-Zvi}, \citenamefont
  {Kamps}, \citenamefont {Liang}, \citenamefont {Muller}, \citenamefont
  {Mueller}, \citenamefont {Padmore}, \citenamefont {Rao}, \citenamefont
  {Tong}, \citenamefont {Vecchione},\ and\ \citenamefont
  {Smedley}}]{KCsSb_APLM2}%
  \BibitemOpen
  \bibfield  {author} {\bibinfo {author} {\bibfnamefont {S.}~\bibnamefont
  {Schubert}}, \bibinfo {author} {\bibfnamefont {M.}~\bibnamefont {Ruiz-Oses}},
  \bibinfo {author} {\bibfnamefont {I.}~\bibnamefont {Ben-Zvi}}, \bibinfo
  {author} {\bibfnamefont {T.}~\bibnamefont {Kamps}}, \bibinfo {author}
  {\bibfnamefont {X.}~\bibnamefont {Liang}}, \bibinfo {author} {\bibfnamefont
  {E.}~\bibnamefont {Muller}}, \bibinfo {author} {\bibfnamefont
  {K.}~\bibnamefont {Mueller}}, \bibinfo {author} {\bibfnamefont
  {H.}~\bibnamefont {Padmore}}, \bibinfo {author} {\bibfnamefont
  {T.}~\bibnamefont {Rao}}, \bibinfo {author} {\bibfnamefont {X.}~\bibnamefont
  {Tong}}, \bibinfo {author} {\bibfnamefont {T.}~\bibnamefont {Vecchione}}, \
  and\ \bibinfo {author} {\bibfnamefont {J.}~\bibnamefont {Smedley}},\
  }\href@noop {} {\bibfield  {journal} {\bibinfo  {journal} {APL-Materials}\
  }\textbf {\bibinfo {volume} {1}},\ \bibinfo {pages} {032119} (\bibinfo {year}
  {2013})}\BibitemShut {NoStop}%
\bibitem [{\citenamefont {Ruiz-Oses}\ \emph {et~al.}(2014)\citenamefont
  {Ruiz-Oses}, \citenamefont {Schubert}, \citenamefont {Attenkofer},
  \citenamefont {Ben-Zvi}, \citenamefont {Liang}, \citenamefont {Muller},
  \citenamefont {Padmore}, \citenamefont {Rao}, \citenamefont {Vecchione},
  \citenamefont {Wong}, \citenamefont {Xie},\ and\ \citenamefont
  {Smedley}}]{KCsSb_APLM}%
  \BibitemOpen
  \bibfield  {author} {\bibinfo {author} {\bibfnamefont {M.}~\bibnamefont
  {Ruiz-Oses}}, \bibinfo {author} {\bibfnamefont {S.}~\bibnamefont {Schubert}},
  \bibinfo {author} {\bibfnamefont {K.}~\bibnamefont {Attenkofer}}, \bibinfo
  {author} {\bibfnamefont {I.}~\bibnamefont {Ben-Zvi}}, \bibinfo {author}
  {\bibfnamefont {X.}~\bibnamefont {Liang}}, \bibinfo {author} {\bibfnamefont
  {E.}~\bibnamefont {Muller}}, \bibinfo {author} {\bibfnamefont
  {H.}~\bibnamefont {Padmore}}, \bibinfo {author} {\bibfnamefont
  {T.}~\bibnamefont {Rao}}, \bibinfo {author} {\bibfnamefont {T.}~\bibnamefont
  {Vecchione}}, \bibinfo {author} {\bibfnamefont {J.}~\bibnamefont {Wong}},
  \bibinfo {author} {\bibfnamefont {J.}~\bibnamefont {Xie}}, \ and\ \bibinfo
  {author} {\bibfnamefont {J.}~\bibnamefont {Smedley}},\ }\href@noop {}
  {\bibfield  {journal} {\bibinfo  {journal} {APL-Materials}\ }\textbf
  {\bibinfo {volume} {2}},\ \bibinfo {pages} {121101} (\bibinfo {year}
  {2014})}\BibitemShut {NoStop}%
\bibitem [{\citenamefont {Karkare}\ and\ \citenamefont
  {Bazarov}(2011)}]{karkare}%
  \BibitemOpen
  \bibfield  {author} {\bibinfo {author} {\bibfnamefont {S.}~\bibnamefont
  {Karkare}}\ and\ \bibinfo {author} {\bibfnamefont {I.}~\bibnamefont
  {Bazarov}},\ }\href@noop {} {\bibfield  {journal} {\bibinfo  {journal} {Appl.
  Phys. Lett.}\ }\textbf {\bibinfo {volume} {98}},\ \bibinfo {pages} {094104}
  (\bibinfo {year} {2011})}\BibitemShut {NoStop}%
\bibitem [{\citenamefont {Bradley}, \citenamefont {Allenson},\ and\
  \citenamefont {Holeman}(1977)}]{bradley}%
  \BibitemOpen
  \bibfield  {author} {\bibinfo {author} {\bibfnamefont {D.~J.}\ \bibnamefont
  {Bradley}}, \bibinfo {author} {\bibfnamefont {M.~B.}\ \bibnamefont
  {Allenson}}, \ and\ \bibinfo {author} {\bibfnamefont {B.~R.}\ \bibnamefont
  {Holeman}},\ }\href@noop {} {\bibfield  {journal} {\bibinfo  {journal} {J.
  Phys. D: Appl. Phys.}\ }\textbf {\bibinfo {volume} {10}},\ \bibinfo {pages}
  {111} (\bibinfo {year} {1977})}\BibitemShut {NoStop}%
\bibitem [{\citenamefont {Smedley}\ \emph {et~al.}(2015)\citenamefont
  {Smedley}, \citenamefont {Gaowei}, \citenamefont {Sinsheimer}, \citenamefont
  {Attenkofer}, \citenamefont {Walsh}, \citenamefont {Schubert}, \citenamefont
  {Wong}, \citenamefont {Padmore}, \citenamefont {Kuhn}, \citenamefont
  {Muller}, \citenamefont {Ding}, \citenamefont {Frisch}, \citenamefont
  {Bhandari}, \citenamefont {Lingertat}, \citenamefont {Wang}, \citenamefont
  {Ovechkina},\ and\ \citenamefont {Nagarkar}}]{JOHN2015_IPAC}%
  \BibitemOpen
  \bibfield  {author} {\bibinfo {author} {\bibfnamefont {J.}~\bibnamefont
  {Smedley}}, \bibinfo {author} {\bibfnamefont {M.}~\bibnamefont {Gaowei}},
  \bibinfo {author} {\bibfnamefont {J.}~\bibnamefont {Sinsheimer}}, \bibinfo
  {author} {\bibfnamefont {K.}~\bibnamefont {Attenkofer}}, \bibinfo {author}
  {\bibfnamefont {J.}~\bibnamefont {Walsh}}, \bibinfo {author} {\bibfnamefont
  {S.}~\bibnamefont {Schubert}}, \bibinfo {author} {\bibfnamefont
  {J.}~\bibnamefont {Wong}}, \bibinfo {author} {\bibfnamefont {H.}~\bibnamefont
  {Padmore}}, \bibinfo {author} {\bibfnamefont {J.}~\bibnamefont {Kuhn}},
  \bibinfo {author} {\bibfnamefont {E.}~\bibnamefont {Muller}}, \bibinfo
  {author} {\bibfnamefont {Z.}~\bibnamefont {Ding}}, \bibinfo {author}
  {\bibfnamefont {H.}~\bibnamefont {Frisch}}, \bibinfo {author} {\bibfnamefont
  {H.~B.}\ \bibnamefont {Bhandari}}, \bibinfo {author} {\bibfnamefont
  {H.}~\bibnamefont {Lingertat}}, \bibinfo {author} {\bibfnamefont
  {V.}~\bibnamefont {Wang}}, \bibinfo {author} {\bibfnamefont {O.}~\bibnamefont
  {Ovechkina}}, \ and\ \bibinfo {author} {\bibfnamefont {V.~V.}\ \bibnamefont
  {Nagarkar}},\ }\href@noop {} {\bibfield  {journal} {\bibinfo  {journal}
  {Proceedings of IPAC 2015}\ ,\ \bibinfo {pages} {TUPHA003}} (\bibinfo {year}
  {2015})}\BibitemShut {NoStop}%
\bibitem [{\citenamefont {Feng}\ \emph {et~al.}(2015)\citenamefont {Feng},
  \citenamefont {Nasiatka}, \citenamefont {Wan}, \citenamefont {Karkare},
  \citenamefont {Smedley},\ and\ \citenamefont {Padmore}}]{jun_thermal}%
  \BibitemOpen
  \bibfield  {author} {\bibinfo {author} {\bibfnamefont {J.}~\bibnamefont
  {Feng}}, \bibinfo {author} {\bibfnamefont {J.}~\bibnamefont {Nasiatka}},
  \bibinfo {author} {\bibfnamefont {W.}~\bibnamefont {Wan}}, \bibinfo {author}
  {\bibfnamefont {S.}~\bibnamefont {Karkare}}, \bibinfo {author} {\bibfnamefont
  {J.}~\bibnamefont {Smedley}}, \ and\ \bibinfo {author} {\bibfnamefont
  {H.}~\bibnamefont {Padmore}},\ }\href@noop {} {\bibfield  {journal} {\bibinfo
   {journal} {Appl. Phys. Lett.}\ }\textbf {\bibinfo {volume} {107}},\ \bibinfo
  {pages} {134101} (\bibinfo {year} {2015})}\BibitemShut {NoStop}%
\bibitem [{\citenamefont {Cultrera}\ \emph {et~al.}(2015)\citenamefont
  {Cultrera}, \citenamefont {Karkare}, \citenamefont {Lee}, \citenamefont
  {Liu}, \citenamefont {Bazarov},\ and\ \citenamefont {Dunham}}]{cold_prstab}%
  \BibitemOpen
  \bibfield  {author} {\bibinfo {author} {\bibfnamefont {L.}~\bibnamefont
  {Cultrera}}, \bibinfo {author} {\bibfnamefont {S.}~\bibnamefont {Karkare}},
  \bibinfo {author} {\bibfnamefont {H.}~\bibnamefont {Lee}}, \bibinfo {author}
  {\bibfnamefont {X.}~\bibnamefont {Liu}}, \bibinfo {author} {\bibfnamefont
  {I.}~\bibnamefont {Bazarov}}, \ and\ \bibinfo {author} {\bibfnamefont
  {B.}~\bibnamefont {Dunham}},\ }\href@noop {} {\bibfield  {journal} {\bibinfo
  {journal} {Phys. Rev. ST - Acc. Beams}\ }\textbf {\bibinfo {volume} {18}},\
  \bibinfo {pages} {113401} (\bibinfo {year} {2015})}\BibitemShut {NoStop}%
\bibitem [{\citenamefont {Sommer}(1963)}]{sommer}%
  \BibitemOpen
  \bibfield  {author} {\bibinfo {author} {\bibfnamefont {A.}~\bibnamefont
  {Sommer}},\ }\href@noop {} {\bibfield  {journal} {\bibinfo  {journal} {Appl.
  Phys. Lett.}\ }\textbf {\bibinfo {volume} {3}},\ \bibinfo {pages} {62}
  (\bibinfo {year} {1963})}\BibitemShut {NoStop}%
\bibitem [{\citenamefont {Kapusta}\ \emph {et~al.}(2007)\citenamefont
  {Kapusta}, \citenamefont {Lavoute}, \citenamefont {Lherbet}, \citenamefont
  {Rossignol}, \citenamefont {Moussant},\ and\ \citenamefont
  {Fouche}}]{alkali_app}%
  \BibitemOpen
  \bibfield  {author} {\bibinfo {author} {\bibfnamefont {M.}~\bibnamefont
  {Kapusta}}, \bibinfo {author} {\bibfnamefont {P.}~\bibnamefont {Lavoute}},
  \bibinfo {author} {\bibfnamefont {F.}~\bibnamefont {Lherbet}}, \bibinfo
  {author} {\bibfnamefont {E.}~\bibnamefont {Rossignol}}, \bibinfo {author}
  {\bibfnamefont {C.}~\bibnamefont {Moussant}}, \ and\ \bibinfo {author}
  {\bibfnamefont {F.}~\bibnamefont {Fouche}},\ }\href@noop {} {\bibfield
  {journal} {\bibinfo  {journal} {IEEE Nuclear Science Symposium Conference
  Record}\ }\textbf {\bibinfo {volume} {1}},\ \bibinfo {pages} {73} (\bibinfo
  {year} {2007})}\BibitemShut {NoStop}%
\bibitem [{\citenamefont {Dowell}, \citenamefont {Bethel},\ and\ \citenamefont
  {Friddell}(1995)}]{dowell_alkali}%
  \BibitemOpen
  \bibfield  {author} {\bibinfo {author} {\bibfnamefont {D.}~\bibnamefont
  {Dowell}}, \bibinfo {author} {\bibfnamefont {S.}~\bibnamefont {Bethel}}, \
  and\ \bibinfo {author} {\bibfnamefont {K.}~\bibnamefont {Friddell}},\
  }\href@noop {} {\bibfield  {journal} {\bibinfo  {journal} {Nucl. Instrum.
  Methods A}\ }\textbf {\bibinfo {volume} {356}},\ \bibinfo {pages} {167}
  (\bibinfo {year} {1995})}\BibitemShut {NoStop}%
\bibitem [{\citenamefont {Nakamura}\ \emph {et~al.}(2010)\citenamefont
  {Nakamura}, \citenamefont {Hamana}, \citenamefont {Ishigami},\ and\
  \citenamefont {Matsui}}]{nakamura}%
  \BibitemOpen
  \bibfield  {author} {\bibinfo {author} {\bibfnamefont {K.}~\bibnamefont
  {Nakamura}}, \bibinfo {author} {\bibfnamefont {Y.}~\bibnamefont {Hamana}},
  \bibinfo {author} {\bibfnamefont {Y.}~\bibnamefont {Ishigami}}, \ and\
  \bibinfo {author} {\bibfnamefont {T.}~\bibnamefont {Matsui}},\ }\href@noop {}
  {\bibfield  {journal} {\bibinfo  {journal} {Nucl. Instrum. Methods A}\
  }\textbf {\bibinfo {volume} {623}},\ \bibinfo {pages} {276} (\bibinfo {year}
  {2010})}\BibitemShut {NoStop}%
\bibitem [{\citenamefont {Schubert}\ \emph {et~al.}(2016)\citenamefont
  {Schubert}, \citenamefont {Wong}, \citenamefont {Feng}, \citenamefont
  {Karkare}, \citenamefont {Padmore}, \citenamefont {Ruiz-Oses}, \citenamefont
  {Smedley}, \citenamefont {Muller}, \citenamefont {Ding}, \citenamefont
  {Gaowei} \emph {et~al.}}]{susanne_PRSTAB}%
  \BibitemOpen
  \bibfield  {author} {\bibinfo {author} {\bibfnamefont {S.}~\bibnamefont
  {Schubert}}, \bibinfo {author} {\bibfnamefont {J.}~\bibnamefont {Wong}},
  \bibinfo {author} {\bibfnamefont {J.}~\bibnamefont {Feng}}, \bibinfo {author}
  {\bibfnamefont {S.}~\bibnamefont {Karkare}}, \bibinfo {author} {\bibfnamefont
  {H.}~\bibnamefont {Padmore}}, \bibinfo {author} {\bibfnamefont
  {M.}~\bibnamefont {Ruiz-Oses}}, \bibinfo {author} {\bibfnamefont
  {J.}~\bibnamefont {Smedley}}, \bibinfo {author} {\bibfnamefont
  {E.}~\bibnamefont {Muller}}, \bibinfo {author} {\bibfnamefont
  {Z.}~\bibnamefont {Ding}}, \bibinfo {author} {\bibfnamefont {M.}~\bibnamefont
  {Gaowei}},  \emph {et~al.},\ }\href@noop {} {\bibfield  {journal} {\bibinfo
  {journal} {J. Appl. Phys.}\ }\textbf {\bibinfo {volume} {In print}} (\bibinfo
  {year} {2016})}\BibitemShut {NoStop}%
\bibitem [{alf()}]{alfaaesar}%
  \BibitemOpen
  \href@noop {} {}\bibinfo {howpublished}
  {\url{https://www.alfa.com/en/}}\BibitemShut {NoStop}%
\bibitem [{SAE()}]{SAES}%
  \BibitemOpen
  \href@noop {} {}\bibinfo {howpublished}
  {\url{https://www.saesgetters.com/}}\BibitemShut {NoStop}%
\bibitem [{\citenamefont {Liang}\ \emph {et~al.}(2012)\citenamefont {Liang},
  \citenamefont {Ruiz-Osés}, \citenamefont {Ben-Zvi}, \citenamefont {Smedley},
  \citenamefont {Attenkofer}, \citenamefont {Vecchione}, \citenamefont
  {Padmore},\ and\ \citenamefont {Schubert}}]{JOHN2012_IPAC}%
  \BibitemOpen
  \bibfield  {author} {\bibinfo {author} {\bibfnamefont {X.}~\bibnamefont
  {Liang}}, \bibinfo {author} {\bibfnamefont {M.}~\bibnamefont {Ruiz-Osés}},
  \bibinfo {author} {\bibfnamefont {I.}~\bibnamefont {Ben-Zvi}}, \bibinfo
  {author} {\bibfnamefont {J.}~\bibnamefont {Smedley}}, \bibinfo {author}
  {\bibfnamefont {K.}~\bibnamefont {Attenkofer}}, \bibinfo {author}
  {\bibfnamefont {T.}~\bibnamefont {Vecchione}}, \bibinfo {author}
  {\bibfnamefont {H.}~\bibnamefont {Padmore}}, \ and\ \bibinfo {author}
  {\bibfnamefont {S.}~\bibnamefont {Schubert}},\ }\href@noop {} {\bibfield
  {journal} {\bibinfo  {journal} {Proceedings of IPAC 2012}\ ,\ \bibinfo
  {pages} {MOPPP049}} (\bibinfo {year} {2012})}\BibitemShut {NoStop}%
\bibitem [{\citenamefont {Czanderna}, \citenamefont {Powell},\ and\
  \citenamefont {Madey}()}]{alkali_sticking}%
  \BibitemOpen
  \bibfield  {author} {\bibinfo {author} {\bibfnamefont {A.~W.}\ \bibnamefont
  {Czanderna}}, \bibinfo {author} {\bibfnamefont {C.~J.}\ \bibnamefont
  {Powell}}, \ and\ \bibinfo {author} {\bibfnamefont {T.~E.}\ \bibnamefont
  {Madey}},\ }\href@noop {} {\emph {\bibinfo {title} {Specimen Handling,
  Preparation, and Treatments in Surface Characterization}}}\ (\bibinfo
  {publisher} {Kluwer Academic Publishers, New York})\BibitemShut {NoStop}%
\bibitem [{\citenamefont {Feng}\ \emph {et~al.}(2013)\citenamefont {Feng},
  \citenamefont {Nasiatka}, \citenamefont {Wong}, \citenamefont {Chen},
  \citenamefont {Hidalgo}, \citenamefont {Vecchione}, \citenamefont {Zhu},
  \citenamefont {Palomares},\ and\ \citenamefont {Padmore}}]{UVmono}%
  \BibitemOpen
  \bibfield  {author} {\bibinfo {author} {\bibfnamefont {J.}~\bibnamefont
  {Feng}}, \bibinfo {author} {\bibfnamefont {J.}~\bibnamefont {Nasiatka}},
  \bibinfo {author} {\bibfnamefont {J.}~\bibnamefont {Wong}}, \bibinfo {author}
  {\bibfnamefont {X.}~\bibnamefont {Chen}}, \bibinfo {author} {\bibfnamefont
  {S.}~\bibnamefont {Hidalgo}}, \bibinfo {author} {\bibfnamefont
  {T.}~\bibnamefont {Vecchione}}, \bibinfo {author} {\bibfnamefont
  {H.}~\bibnamefont {Zhu}}, \bibinfo {author} {\bibfnamefont {F.~J.}\
  \bibnamefont {Palomares}}, \ and\ \bibinfo {author} {\bibfnamefont {H.~A.}\
  \bibnamefont {Padmore}},\ }\href@noop {} {\bibfield  {journal} {\bibinfo
  {journal} {Rev. Sci. Instrum.}\ }\textbf {\bibinfo {volume} {84}},\ \bibinfo
  {pages} {085114} (\bibinfo {year} {2013})}\BibitemShut {NoStop}%
\bibitem [{\citenamefont {Gorlov}(2007)}]{gorlov}%
  \BibitemOpen
  \bibfield  {author} {\bibinfo {author} {\bibfnamefont {T.}~\bibnamefont
  {Gorlov}},\ }\href@noop {} {\bibfield  {journal} {\bibinfo  {journal}
  {Journal of Electrostatics}\ }\textbf {\bibinfo {volume} {65}},\ \bibinfo
  {pages} {735} (\bibinfo {year} {2007})}\BibitemShut {NoStop}%
\bibitem [{\citenamefont {Sannibale}\ \emph {et~al.}(2016)\citenamefont
  {Sannibale}, \citenamefont {Doyle}, \citenamefont {Feng}, \citenamefont
  {Filippetto}, \citenamefont {Gierman}, \citenamefont {Harris}, \citenamefont
  {Johnson}, \citenamefont {Kramasz}, \citenamefont {Leitner}, \citenamefont
  {Li} \emph {et~al.}}]{APEX}%
  \BibitemOpen
  \bibfield  {author} {\bibinfo {author} {\bibfnamefont {F.}~\bibnamefont
  {Sannibale}}, \bibinfo {author} {\bibfnamefont {J.}~\bibnamefont {Doyle}},
  \bibinfo {author} {\bibfnamefont {J.}~\bibnamefont {Feng}}, \bibinfo {author}
  {\bibfnamefont {D.}~\bibnamefont {Filippetto}}, \bibinfo {author}
  {\bibfnamefont {S.}~\bibnamefont {Gierman}}, \bibinfo {author} {\bibfnamefont
  {G.}~\bibnamefont {Harris}}, \bibinfo {author} {\bibfnamefont
  {M.}~\bibnamefont {Johnson}}, \bibinfo {author} {\bibfnamefont
  {T.}~\bibnamefont {Kramasz}}, \bibinfo {author} {\bibfnamefont
  {D.}~\bibnamefont {Leitner}}, \bibinfo {author} {\bibfnamefont
  {R.}~\bibnamefont {Li}},  \emph {et~al.},\ }\href@noop {} {\bibfield
  {journal} {\bibinfo  {journal} {Proceedings of IPAC 2016}\ ,\ \bibinfo
  {pages} {TUOCA02}} (\bibinfo {year} {2016})}\BibitemShut {NoStop}%
\bibitem [{\citenamefont {Karkare}\ and\ \citenamefont
  {Bazarov}(2015)}]{nonuniformity}%
  \BibitemOpen
  \bibfield  {author} {\bibinfo {author} {\bibfnamefont {S.}~\bibnamefont
  {Karkare}}\ and\ \bibinfo {author} {\bibfnamefont {I.}~\bibnamefont
  {Bazarov}},\ }\href@noop {} {\bibfield  {journal} {\bibinfo  {journal} {Phys.
  Rev. Applied}\ }\textbf {\bibinfo {volume} {4}},\ \bibinfo {pages} {024015}
  (\bibinfo {year} {2015})}\BibitemShut {NoStop}%
\end{thebibliography}
\end{document}